\documentclass[11pt]{article}

\usepackage[final]{acl}
\usepackage{booktabs}
\usepackage{times}
\usepackage{latexsym}

\usepackage[T1]{fontenc}
\usepackage[utf8]{inputenc}
\usepackage{listings}
\usepackage{microtype}

\usepackage{inconsolata}

\usepackage{graphicx}

\title{Real-Time World Crafting: Generating Structured Game Behaviors from Natural Language with Large Language Models}

\author{Austin Drake\thanks{Corresponding Author} \\
  University of Exeter \\
  Department of Computer Science \\
  \texttt{austin.drake.343@gmail.com} \\\And
  Hang Dong \\
  University of Exeter \\
  Department of Computer Science \\
  \texttt{h.dong2@exeter.ac.uk} \\}

\begin{document}
\maketitle
\begin{abstract}
We present a novel architecture for safely integrating Large Language Models (LLMs) into interactive game engines, allowing players to ``program'' new behaviors using natural language. Our framework mitigates risks by using an LLM to translate commands into a constrained Domain-Specific Language (DSL), which configures a custom Entity-Component-System (ECS) at runtime. We evaluated this system in a 2D spell-crafting game prototype by experimentally assessing models from the \textit{Gemini}, \textit{GPT}, and \textit{Claude} families with various prompting strategies. A validated LLM judge qualitatively rated the outputs, showing that while larger models better captured creative intent, the optimal prompting strategy is task-dependent: Chain-of-Thought improved creative alignment, while few-shot examples were necessary to generate more complex DSL scripts. This work offers a validated LLM-ECS pattern for emergent gameplay and a quantitative performance comparison for developers.
\end{abstract}

\section{Introduction}

\begin{figure*}[t]
\centering	
\includegraphics[width=\linewidth]{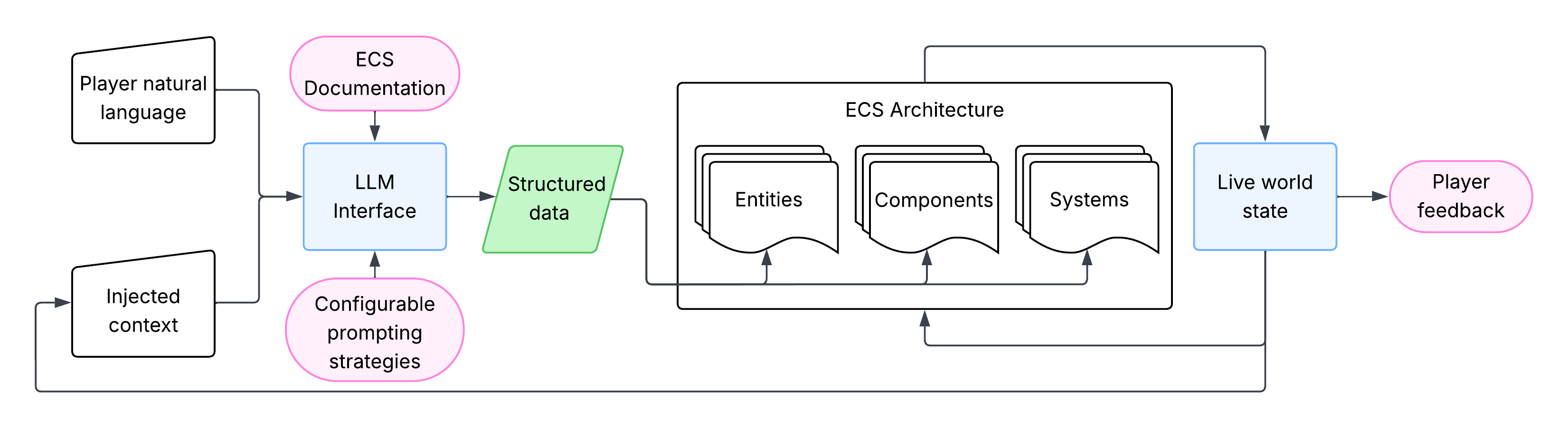}
\caption{A flow diagram of the developed architecture. The LLM interface (left) handles translating Natural Language (NL) into DSL code. The ECS (center) consists of data output by the LLM and logical systems operating on that data. The I/O container (outside edges) is the mode of content delivery and interaction for the player.
}\label{fig:arch_diagram}
\end{figure*}

A long-standing goal in Human-Computer Interaction (HCI) and in interactive art forms is to allow users to control complex simulations in real-time using only natural language, which requires translating between ambiguous human intent and the rigid logic of computer systems \cite{nlp-for-hci}. While Large Language Models (LLMs) offer a promising approach to this problem, allowing an LLM to generate and execute arbitrary code in a general-purpose language poses significant risks to system stability and security \cite{majumdar2024towards}. To mitigate these risks, we propose a novel software architecture that uses an LLM to interpret natural language commands into a constrained, human-readable Domain-Specific Language (DSL). This intermediate layer provides a safe and verifiable instruction set for a video game simulation, addressing a gap in research on the real-time translation of creative instructions and producing physically plausible behaviors in interactive systems. We use a 2D physics-based video game as a testbed to evaluate this architecture.

While many games use LLMs for narrative generation, our work focuses on driving more realized, physically-based systems. Our proposed architecture relies on the Entity-Component-System (ECS) pattern \cite{Romeo2016, ECSforRIS}, whose flexible, compositional design is well-suited for dynamically modifying object behaviors at runtime based on non-deterministic LLM outputs. Our prototype, \textit{Latent Space}, features a spell-crafting system and a cellular automata sandbox to test the LLM's ability to generate both compositional and logic-based DSL code.

This paper investigates how effectively LLMs can translate creative commands into structured, executable instructions. We experimentally evaluate the performance of models from the \textit{Gemini, GPT,} and \textit{Claude} families with various prompt engineering strategies to determine which factors most influence the generation of logically consistent and physically plausible game mechanics. Our contributions include: (1) a validated, DSL-mediated architecture for safely integrating LLMs into interactive systems; (2) a quantitative comparison of model performance on creative and logical generation tasks; (3) an analysis of prompting strategies for maximizing creative throughput in language-driven interfaces; and (4) an evaluation of the architecture, including both human and automated LLM-as-judge ratings. Our demo, experiment implementation, and generated data are available as a GitHub repository\footnote{\url{https://github.com/austin-the-drake/real-time-world-crafting-wordplay-demo}}.

\section{Background}
\subsection{Historical Challenges in Games}

Natural language interfaces in interactive systems and art forms have historically been limited by the challenges of linguistic ambiguity and the prohibitive authoring cost required to support wide-reaching player freedom. Early parser-based systems such as \textit{Zork} \cite{zorkFeature1979}, while revolutionary, were also very rigid, and frustratingly rejected any inputs that did not exactly match their pre-programmed vocabularies. Later NLP-driven experiences including \textit{Façade} \cite{MateasSternGDC2003, MateasSternAIIDE2005} addressed this problem by mapping free-form user input to the closest available pre-written options, but this comes with an immense workload of manually authoring responses to maintain the illusion of natural conversation. These challenges have led modern games to often rely on systems such as ``dialogue wheel'', which restrict player input to lessen the authoring burden at the cost of player expressiveness \cite{Taylor-Giles2020}.

\begin{figure*}[t]
\centering

\begin{minipage}{0.49\textwidth}
\includegraphics[width=\linewidth]{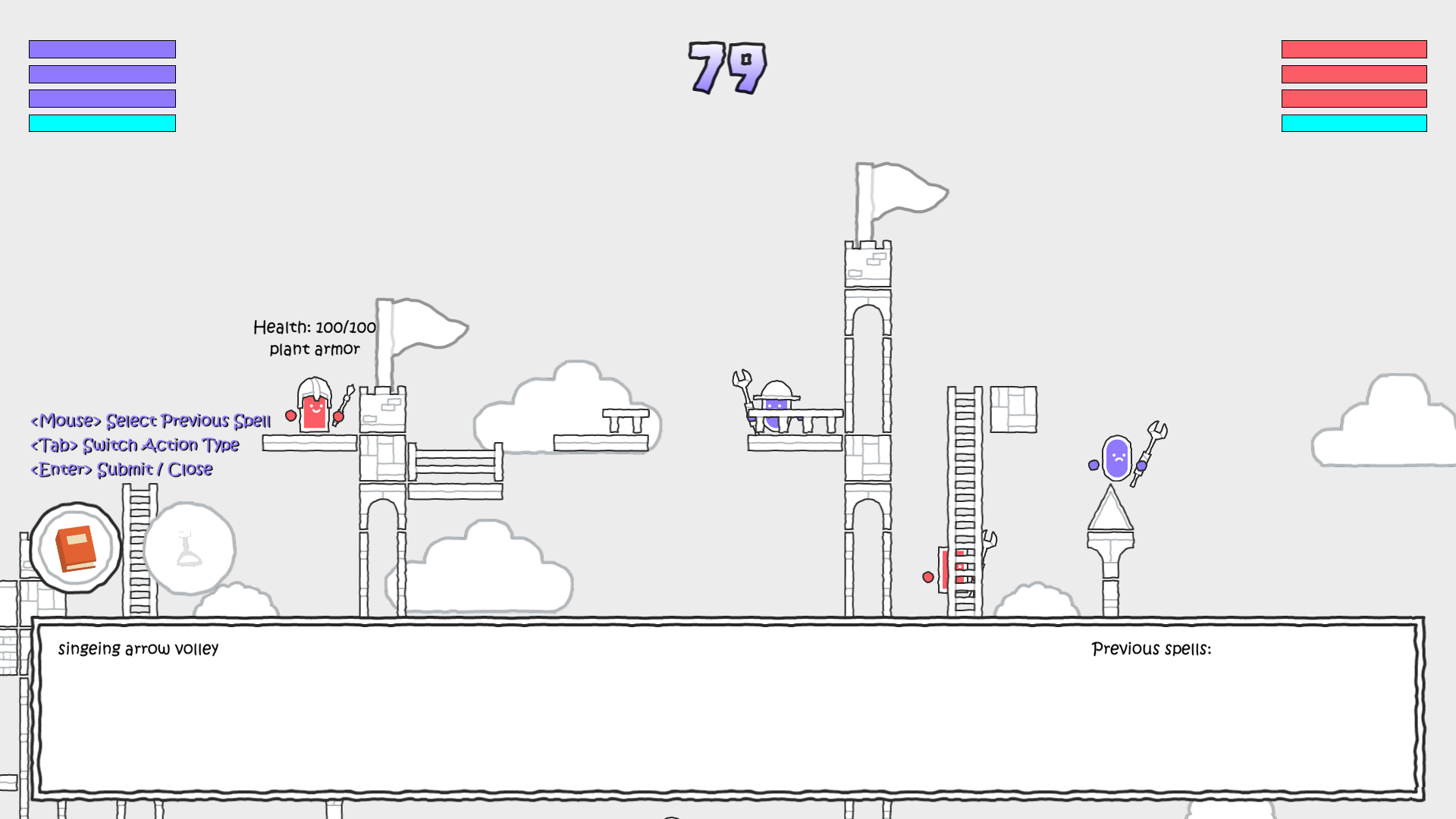}
\end{minipage}
\hfill
\begin{minipage}{0.49\textwidth}
\includegraphics[width=\linewidth]{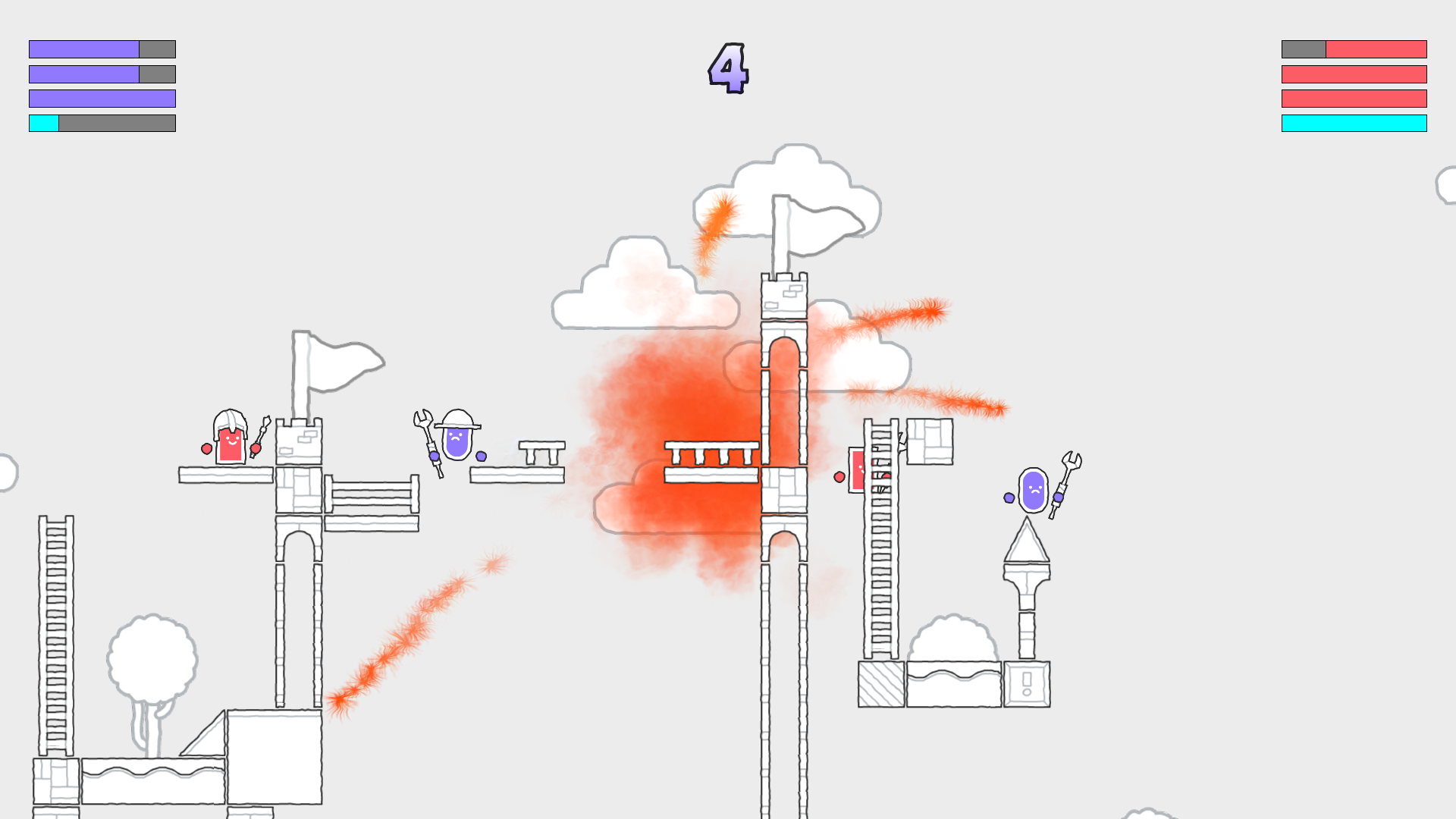}
\end{minipage}

\caption{A pair of screenshots from \textit{Latent Space's} Battle Mode. Natural language magic descriptions, e.g., ``Singeing arrow volley,'' entered into the text interface (left), can have explosive consequences (right). Each team's remaining players and their health (red/purple) are visible above, as is the amount of time remaining for each turn.}
\label{fig:spell_screens}
\end{figure*}

\subsection{LLM-Driven Architectures}

LLMs can simultaneously interpret ambiguous text and generate novel content, which presents a potential solution to both historical problems at once. Contemporary research is moving from using models to select from pre-written options (as in \textit{Façade}) towards generating fully dynamic outcomes in real-time \cite{LLMsInGamesSurvey}. Applications range from generating dynamic narratives in co-creative storytelling and role-playing games \cite{sun2023languagerealitycocreativestorytelling, JazzVsWafflesThesis, aiDungeonMaster} to driving complex agent behaviors in simulations and strategy games \cite{park2023generativeagentsinteractivesimulacra, yim2024evaluatingenhancingllmsagent, behaviorBranch2024}. A promising direction for LLMs is to support natural language as a free-form scripting interface for game mechanics. Several existing projects constrain player input to make this task more tractable \cite{WizardCats}; However, this project investigates the challenge of translating fully unconstrained language into game logic.

Architectural inspiration for this task can be drawn from domains like robotics and data science \cite{naturalLanguageRoboticsReview2024, li2025interactivetaskplanninglanguage, sah2024nl4dvllm}. A common pattern involves an LLM translating a high-level command into a structured, machine-readable representation (e.g., API calls, JSON), which is then executed by a deterministic backend \cite{PERRAULT1988133, sah2024nl4dvllm}. Robotics frameworks like Interactive Task Planning (ITP) use this to break abstract goals down into concrete actions or update plans based on new context \cite{li2025interactivetaskplanninglanguage}. A major focus is grounding the LLM's output in the state and capabilities of the target system \cite{groundedInteractionInGamesWorkshop}.

%\begin{figure}[h!]

%\caption{An example of the magical spell scripting DSL (left) and the cellular automata DSL (right). In natural language, these are A controllable wind pixie that warps me when I call,'' and a cloudy gas that diffuses randomly,'' respectively.}
%\label{fig:dsl_examples}
%\end{figure}

\section{System Architecture and Implementation}

Our demo, \textit{Latent Space}, is designed to translate natural language into executable game mechanics. The primary design challenge was to safely and reliably use the non-deterministic output of an LLM within a deterministic game engine. To achieve this, we developed a modular architecture (Figure \ref{fig:arch_diagram}) that sandboxes the LLM's influence. Instead of generating general-purpose code, which is unconstrained and error-prone, the LLM's role is strictly limited to generating structured data in a custom JSON-based DSL.

The architecture's layers are: (1) an LLM interface that translates a player's request into the DSL; (2) a custom ECS framework that parses the DSL to instantiate and configure the corresponding game entities and their components; and (3) a commercial game engine that acts as an I/O container, rendering the ECS state and capturing player input. This separation of concerns improves system stability and also allows the LLM's translation performance to be tested in isolation.

\subsection{Prototype and DSL Design}
\textit{Latent Space} is designed to provide players with unconstrained, free-form natural language input, in contrast to systems that rely on combining predefined elements \cite{WizardCats}. The prototype features two distinct game modes to evaluate the LLM's reasoning on different types of tasks. Each mode requires its own DSL; both of which are represented as JSON, behaving like an abstract grammar \cite{wang2023grammar} that simplifies parsing.

\textbf{Battle Mode} (Figure \ref{fig:spell_screens}) is a turn-based artillery game where players interact by simply describing magical spells. This mode tests the LLM's ability to perform creative, semantic translation, mapping a user's description to a set of components in our compositional spell DSL. This DSL defines spells as unordered collections of data components (e.g., projectile, element) and event triggers that can embed other spell definitions.

\begin{figure}[t]
\centering
\includegraphics[width=\linewidth]{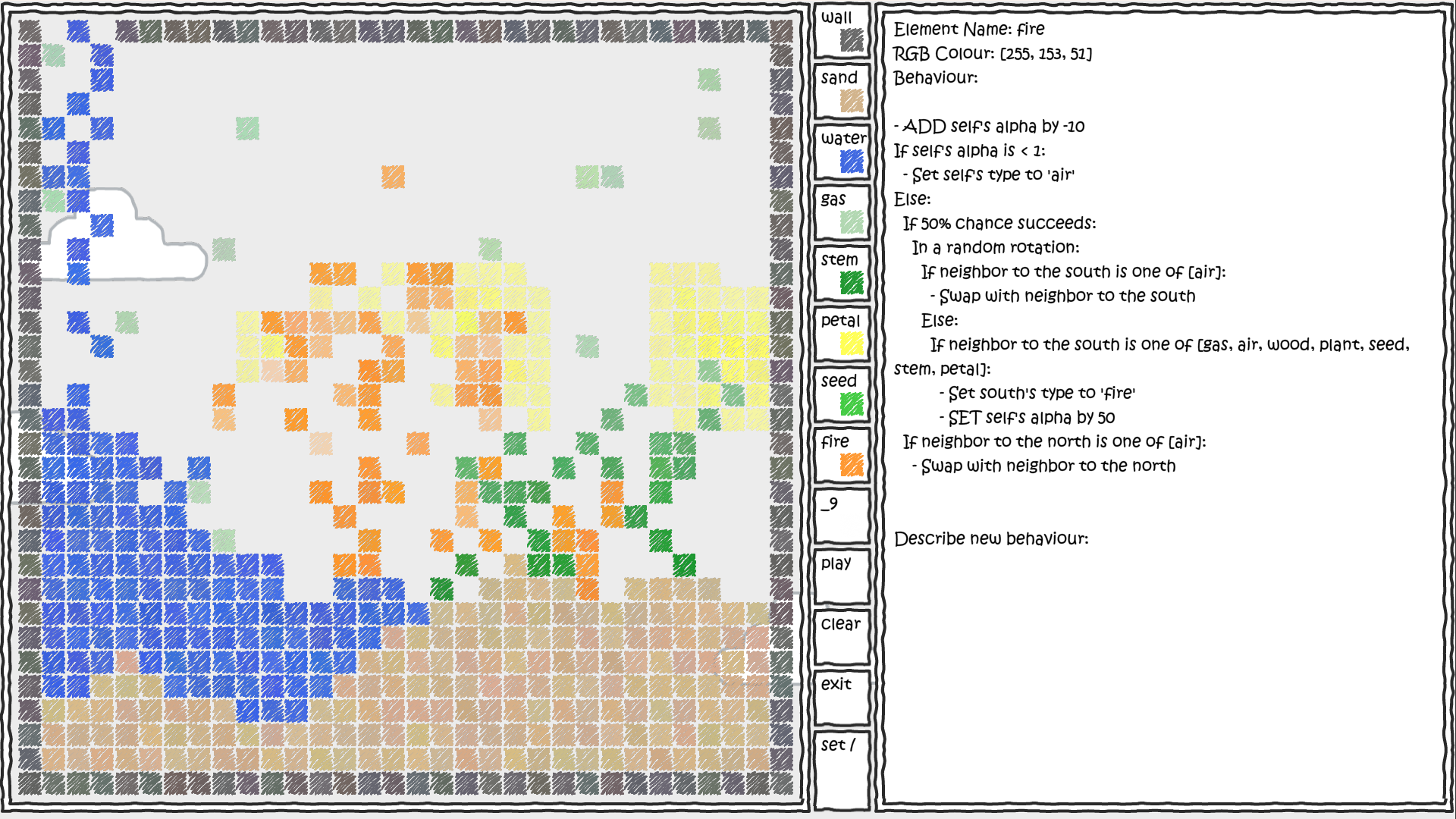}
\caption{A screenshot from \textit{Latent Space's} Alchemy Mode. Natural language material descriptions are translated into actionable cellular automata rules (right) by the LLM. Players can observe the generated behaviors in the workspace (left).}
\label{fig:other_screens}
\end{figure}

\textbf{Alchemy Mode} (Figure \ref{fig:other_screens}) is a cellular automata sandbox modeled after \textit{The Powder Toy} \cite{ThePowderToy} and \textit{Sandspiel} \cite{sandspiel_studio}. This mode tests the LLM's ability to generate complex, ordered logical rules. Players describe new materials, and the LLM must generate a valid ruleset in our procedural cellular automata DSL. This DSL uses a nested structure of conditional checks to define grid-based interactions where the order of operations is critical.

\subsection{Grounding and Robustness}
To ensure system stability, we utilize a DSL as an intermediate layer, which constrains the LLM's output to a set of pre-validated operations, reducing the risks associated with arbitrary code generation \cite{majumdar2024towards}. Since a pre-trained LLM has no knowledge of a novel DSL, this presents an out-of-distribution problem \cite{OOD2023}, which can lead to poor performance and hallucinated function calls \cite{joel2024surveyllmbasedcodegeneration}. Rather than extensive fine-tuning, we use prompt-based knowledge injection, providing the DSL documentation to the model in prompts to ground its output, applying recommendations from literature on DSL code generation \cite{gu2025effectivenesslargelanguagemodels}.

We employ several prompt engineering techniques to ensure the generated DSL code is syntactically and logically sound. To ground the LLM's output in the current game state, each API request is extended with dynamic context. For the spell DSL, this includes a list of currently active magical elements, while for the more complex automata DSL, the prompt includes the full DSL representation of all existing materials. We also use few-shot examples and Chain-of-Thought (CoT) prompting, where the LLM first generates a plan mapping user phrases to DSL components before producing the final JSON \cite{originalCoT, CotForCoding, li-etal-2024-simple}. Our system's backend follows the ECS pattern, a compositional design where an object's behavior is defined by its data components rather than a fixed class hierarchy \cite{Romeo2016, 10043019}. In this model, an Entity is a unique identifier for an object, which holds various Components, consisting of raw data chunks such as position or health. Systems then contain the logic, iterating over entities that possess a specific set of components to apply behaviors such as movement. This separation of data and logic provides the runtime flexibility necessary to dynamically alter game objects by modifying their data components, which is done simply by following the instructions contained in the generated DSL scripts, making it ideal for encouraging emergent gameplay \cite{ECSforRIS, vico2021}. The prompts for spell and automata scripting are available in Appendix \ref{app:spell-prompt} and \ref{app:automata-prompt}, respectively. 

To handle the non-deterministic nature of LLMs, the system also includes several error-handling measures. A validation layer checks all generated JSON for syntactic correctness, valid component types, and in-range parameter values. The parser strips extraneous text, and for minor errors such as a missing parameter, the system applies default values. If a script is completely unusable, the game defaults to a harmless ``fizzle'' effect, providing feedback to the player without crashing the game.

\section{Experimental Methodology}

We evaluated our framework through two primary experiments, using a combination of quantitative algorithmic metrics and automated qualitative ratings from a validated LLM judge. To support this, we generated a corpus of 2600 DSL scripts comprised of three distinct subsets, described below.

\paragraph{Test Data Generation}
First, a naturalistic set (N=2400) was created to test the models' ability to reason with novel concepts that may not have a perfect correspondence to the available DSL components, as would be expected during real gameplay. This set was generated by prompting the \textit{Gemini 2.5 Pro} model \cite{comanici2025gemini25pushingfrontier} to create 100 unique, creative task descriptions, which were then used as inputs for the main experiment. Secondly, a bidirectional set (N=120) was created to test information preservation. It consists of 30 random, procedurally-generated ``source'' scripts and 120 corresponding natural language descriptions of varying styles (a creative narrative vs. a technical document) and lengths (summary length vs detailed report), which were also generated by prompting the \textit{Gemini 2.5 Pro} model. Finally, a handcrafted ground-truth set (N=80) of expert-authored ``good'' (well-formed and effective) and ``bad'' (syntactically valid but logically flawed) scripts was created to rigorously validate the LLM judge. The latter scripts were created by randomly replacing components or by shortening the former. These consisted of 50 spell and 30 cellular automata examples, for a total of 25 ``good'' and 25 ``bad'' spell scripts, and 15 ``good'' and 15 ``bad'' automata scripts.

\subsection{Experiment 1: NL-to-DSL Translation}
This experiment assessed the main translation pipeline using the naturalistic test set in a 4×3×2 fully-crossed, repeated-measures design. We tested three factors: Model (\textit{Gemini 2.5 Flash, GPT-4.1 mini, Claude 4 Sonnet}, and \textit{Gemma 3 4B} as a baseline); Shot Strategy (zero-, one-, and few-shot); and Prompting Technique (standard vs. Chain-of-Thought). Outcome measures included the Average Success Rate (ASR) of generating syntactically valid DSL and, for all valid scripts, automated qualitative ratings from our LLM judge.

\begin{figure*}[t]
\begin{minipage}[t]{0.48\textwidth}
\begin{lstlisting}[label={lst:spell}, captionpos=b, breaklines=true, basicstyle=\small\ttfamily, frame=single]
{
friendlyName: "Wind scout",
count: 1,
components: [
{componentType: "projectile",
    radius: 2, speed: 15, gravity: 0},
{componentType: "element",
    element: "wind"},
{componentType: "controllable",
    mana_cost: 0.1},
{componentType: "buttonTrigger",
    payload_components: [
    {componentType: "teleportCaster"}
  ]}
]}
\end{lstlisting}
\end{minipage}
\hfill
\begin{minipage}[t]{0.48\textwidth}
\begin{lstlisting}[label={lst:ca}, captionpos=b, breaklines=true, basicstyle=\small\ttfamily, frame=single]
{
name: "gas",
color_hex: "#CCCCCC",
behavior: {
  actions: [{
    type: "in_rand_rotation",
    actions: [{
      type: "if_neighbor_is",
      direction: "south",
      options: ["air"],
      actions: [{
        type: "do_swap",
        direction: "south"
        }],
...
\end{lstlisting}
\end{minipage}
\caption{An example of the magical spell scripting DSL (left) and the cellular automata DSL (right). In natural language, these are ``A controllable wind pixie that warps me when I call,'' and ``a cloudy gas that diffuses randomly.''}
\label{fig:dsl_examples}
\end{figure*}

\subsection{Experiment 2: Bidirectional Translation}
This experiment evaluated how well a user's creative intent is preserved in a ``round-trip'' translation, simulating the process of a player articulating a mental concept. Using the bidirectional set, an LLM had to recreate an original, procedurally-generated DSL script using only its corresponding natural language description. Description length and style were compared in a 2x2 factorial design. We assessed the difference between the original and final scripts using two algorithmic similarity measures: (1) tree edit distance, computed between the Abstract Syntax Trees (ASTs) of the scripts using the APTED algorithm to quantify structural changes \cite{PAWLIK2016157, 10.1145/2699485}, and (2) Jaccard similarity, calculated as the intersection-over-union of component names in the scripts, to measure the overlap in pure semantic content. This experiment was limited to the spell DSL.

\subsection{Automated Qualitative Assessment}
Because large-scale human evaluation was infeasible, we used an automated LLM judge based on \textit{GPT 4.1} to provide qualitative ratings at scale, justifying its use with a rigorous validation process. For all valid outputs, the judge assigned 1-5 Likert scores for the 4 criteria described below.

% \begin{itemize}
%     \item \textbf{Creative Alignment} assesses how well the generated output captured the user's core creative intent, with an emphasis on theme.
%     \item \textbf{Instruction Following} measures how accurately the output adhered to specific, direct commands or constraints in player requests.
%     \item \textbf{Emergence} measures whether the output surpassed the user's request in a positive and unexpected way, such as environmental interaction.
%     \item \textbf{Structural Coherence} evaluates adherence to the DSL's logical rules. This was designed to provide a more comprehensive measure after basic syntactic validity was found to be very high.
% \end{itemize}
\paragraph{(i) Creative Alignment:} This scale assesses how well the generated output captured the user's core creative intent, with an emphasis on theme.

\paragraph{(ii) Instruction Following:} This scale measures how accurately the output adhered to specific, direct commands or constraints in player requests.

\paragraph{(iii) Emergence:} Measures whether the output surpassed the user's request in a positive and unexpected way, such as environmental interaction.

\paragraph{(iv) Structural Coherence:} Evaluates adherence to the DSL's logical rules. This was designed to provide a more comprehensive measure after basic syntactic validity was found to be very high.
\\[\baselineskip]
The LLM judge (\textit{GPT 4.1}) was instructed to produce a textual evaluation of each script before assigning a numerical score, a form of Chain-of-Thought prompting used to improve reliability \cite{originalCoT}. To protect against a common leniency bias in LLM judges, a large model and careful prompting were used in accordance with established guidelines \cite{thakur2025judgingjudgesevaluatingalignment}.
The prompt for the LLM judge is available in Appendix \ref{app:judge-prompt}.

\begin{table*}[t]
\centering
\fontsize{9}{11}\selectfont
\caption{Percentage of Success (ASR, score in \%) Across Models and Prompting Strategies}
\label{tab:asr_results}
\begin{tabular*}{\textwidth}{@{\extracolsep{\fill}}lcccccc@{}}
\toprule
& \multicolumn{2}{c}{\textbf{Zero-Shot}} & \multicolumn{2}{c}{\textbf{One-Shot}} & \multicolumn{2}{c}{\textbf{Few-Shot}} \\
\cmidrule(lr){2-3} \cmidrule(lr){4-5} \cmidrule(lr){6-7}
\textbf{Model} & \textbf{Standard} & \textbf{CoT} & \textbf{Standard} & \textbf{CoT} & \textbf{Standard} & \textbf{CoT} \\
\midrule
\multicolumn{7}{l}{\textbf{Magical Spells}} \\

Claude 4 Sonnet   & 100 & 100 & 100 & 100 & 100 & 100 \\
Gemini 2.5 Flash  & \textbf{96} & 100 & 100 & \textbf{98} & 100 & 100 \\
Gemma 3 (4B)      & 100 & \textbf{98} & 100 & 100 & 100 & 100 \\
GPT-4.1 Mini      & 100 & 100 & 100 & 100 & 100 & 100 \\

\multicolumn{7}{l}{\textbf{Cellular Automata}} \\

Claude 4 Sonnet   & 100 & 100 & 100 & 100 & \textbf{98} & 100 \\
Gemini 2.5 Flash  & \textbf{82} & \textbf{92} & 100 & 100 & \textbf{94} & \textbf{98} \\
Gemma 3 (4B)      & \textbf{92} & \textbf{76} & \textbf{98} & 100 & 100 & 100 \\
GPT-4.1 Mini      & \textbf{90} & \textbf{88} & \textbf{98} & \textbf{98} & 100 & 100 \\
\bottomrule
\end{tabular*}
\end{table*}

\paragraph{Validating the LLM judge} The judge was validated against the handcrafted ground-truth dataset (N=80) as introduced earlier in the Section. A series of paired Wilcoxon Signed-Rank tests confirmed it could reliably distinguish between ``good'' and ``bad'' examples. Scores for ``good'' items were statistically significantly higher across all criteria: Creative Alignment (V=147, p=0.00058), Instruction Following (V=166, p=0.00037), Emergence (V=271, p<.0001), and Structural Coherence (V=195.5, p<.00057). The judge also had strong classification performance; for the 50 spell scripts, maximum F1 scores were 0.77, 0.81, 0.90, and 0.80, and Area Under the Curve (AUC) values were 0.82, 0.85, 0.92, and 0.82, respectively, on the four criteria. For automata scripts, F1 scores were 0.83, 0.93, 0.80, and 0.77, and AUC values were 0.88, 0.96, 0.83, and 0.70. This accuracy demonstrates alignment with a human-defined quality standard.

To test for consistency, a second, equivalently-capable LLM (\textit{Gemini 2.5 Pro}) rated the same outputs to calculate inter-rater reliability using Spearman's rank-order correlation ($\rho$), quadratically weighted Cohen's Kappa ($\kappa_w$), and a two-way Intraclass Correlation Coefficient (ICC). Agreement was moderate to substantial; For magical spells, Spearman's $\rho$ ranged from .59 to .70 across scales, $\kappa_w$ from .56 to .64, and ICC from .56 to .65. For cellular automata, agreement was higher ($\rho$ from .76 to .81, $\kappa_w$ from .74 to .77, and ICC from .75 to .77), with the exception of Structural Coherence, which showed lower agreement ($\rho$=.36, $\kappa_w$=.23, ICC=.24). The full metrics are available in Appendix B (Table \ref{tab:irr_metrics}). Substantial agreement indicates the criteria are objective enough to be scored consistently, reducing the concern of single-model bias. However, we acknowledge self-enhancement bias \cite{zheng2023judgingllmasajudgemtbenchchatbot} as a potential limitation and revisit it in the discussion section.

\section{Experimental Results}

The overall reliability of the translation pipeline, measured by Average Success Rate (ASR), was high. For the compositional spell DSL, success was nearly perfect, whereas the procedural automata DSL proved slightly more challenging (Table \ref{tab:asr_results}). This high syntactic validity can be attributed to both the maturity of modern LLMs and the pipeline's error-handling features, which salvaged most flawed outputs. However, this large ceiling effect means that a simple pass/fail metric is limited in utility compared to the qualitative measures.

\subsection{NL-to-DSL Translation}

An ANOVA revealed that model choice was the strongest predictor of performance (p<.001 on all outcomes), as shown in Table \ref{tab:exp1_anova} in Appendix A. Post-hoc tests showed that for the creative spell task, \textit{Claude 4 Sonnet} significantly outperformed all other models on the automated qualitative ratings. For the logical automata task, this gap narrowed, but \textit{Claude 4 Sonnet} still significantly outperformed other models on most outcomes. The small baseline model, \textit{Gemma 3 (4B)}, performed significantly worse than all larger models across both tasks. The effect of prompting strategy was also task-dependent. For the compositional spell task, in-context examples had no significant effect. For the procedural automata task, however, few-shot and one-shot prompting significantly improved all scores over zero-shot. Chain-of-Thought (CoT) prompting was a more generally effective strategy, significantly improving Creative Alignment and Emergence for both DSLs. We also observed several significant interaction effects (Figure \ref{fig:exp1_interactions} in Appendix A), showing that the most effective strategy is highly model-dependent.

\subsection{Bidirectional Translation}
\begin{table*}[t]
\centering
\footnotesize
\caption{Linear Mixed-Model Results for the bidirectional translation experiment}
\label{tab:exp2_lmm}
\begin{tabular*}{\textwidth}{@{\extracolsep{\fill}}lcccccc@{}}
\toprule
& \multicolumn{3}{c}{\textbf{Tree Edit Distance}} & \multicolumn{3}{c}{\textbf{Jaccard Similarity}} \\
\cmidrule(lr){2-4} \cmidrule(lr){5-7}
\textbf{Predictor} & \textbf{Estimate} & \textbf{SE} & \textbf{p} & \textbf{Estimate} & \textbf{SE} & \textbf{p} \\
\midrule
\multicolumn{7}{l}{\textit{Fixed Effects}} \\
(Intercept) & 4.34 & 2.19 & .056 & \textbf{0.36} & \textbf{0.05} & \textbf{$<$.001\ddag} \\
Description Style (Technical) & \textbf{-7.40} & \textbf{1.10} & \textbf{$<$.001\ddag} & \textbf{0.13} & \textbf{0.02} & \textbf{$<$.001\ddag} \\
Description Detail (Summary) & 0.87 & 1.10 & .433 & \textbf{-0.11} & \textbf{0.02} & \textbf{$<$.001\ddag} \\
Component Complexity & -0.31 & 0.44 & .485 & -0.00 & 0.01 & .854 \\
Nesting Complexity & \textbf{1.86} & \textbf{0.45} & \textbf{$<$.001\ddag} & 0.02 & 0.01 & .096 \\
Style $\times$ Detail Interaction & \textbf{5.60} & \textbf{1.56} & \textbf{$<$.001\ddag} & -0.03 & 0.03 & .305 \\
\addlinespace
\multicolumn{7}{l}{\textit{Random Effects}} \\
& \textbf{Variance} & \textbf{Std. Dev.} & & \textbf{Variance} & \textbf{Std. Dev.} & \\
\cmidrule(lr){2-3} \cmidrule(lr){5-6}
Procedural Code (Intercept) & 6.85 & 2.62 & & 0.005 & 0.070 & \\
Residual & 18.15 & 4.26 & & 0.009 & 0.092 & \\
\bottomrule
\multicolumn{7}{l}{\textit{Note.} Significance codes: * $p < .05$, \dag\ $p < .01$, \ddag\ $p < .001$. Intercept contains Style (Narrative) and Detail (Detailed)} \\
\end{tabular*}
\end{table*}

To assess how well a user's creative intent is preserved, this experiment performed a round-trip translation from a DSL script to natural language and back again. We specifically investigated how the style (narrative vs. technical) and length (summary vs. detailed) of the natural language description impacted the accuracy of the final, regenerated script, using a linear mixed-effect model to handle the mixture of discrete and continuous variables and 2x2 factorial design.

\begin{figure}[h!]
\centering	
\includegraphics[width=\linewidth]{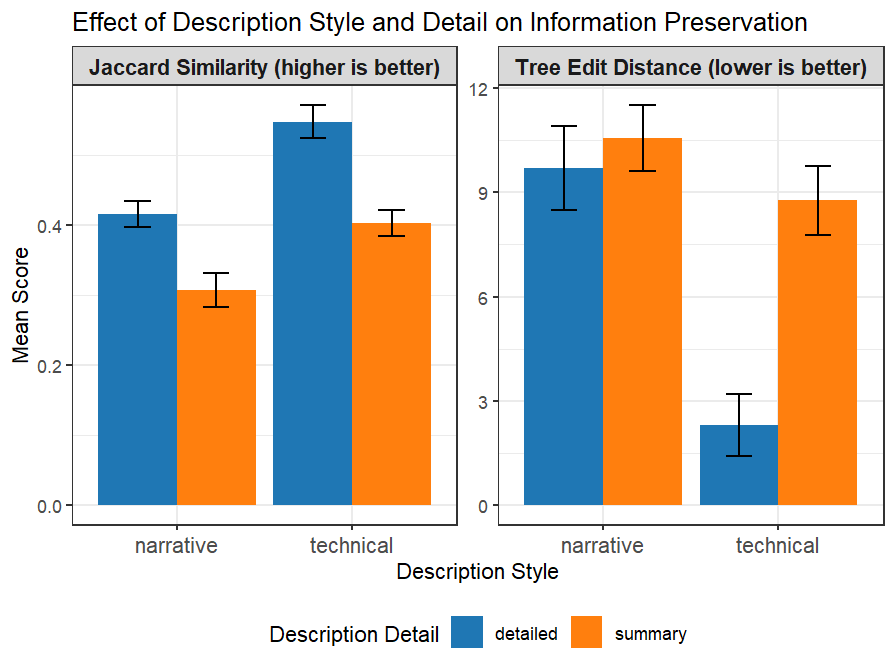}
\caption{Clustered bar chart for the bidirectional translation experiment. Left: Jaccard Similarity (higher is better). Right: Tree Edit Distance (lower is better).
}\label{fig:exp2_bars}
\end{figure}

The linear mixed-effect model results for the bidirectional experiment are shown in Table \ref{tab:exp2_lmm}, and visualized in Figure \ref{fig:exp2_bars}. The most influential factor was the style of the natural language description; Technical descriptions resulted in significantly better information preservation than creative narrative descriptions, with scripts that were both more structurally similar (lower tree edit distance, $p <.001$) and had greater semantic overlap (higher Jaccard Similarity, $p <.001$) with the originals. The complexity of the original script was also a significant factor; scripts with greater nesting complexity were harder to reproduce, resulting in a higher tree edit distance (p<.001). The length of the description was observed to influence semantic content, as short descriptions resulted in significantly lower Jaccard Similarity (p <.001). A significant interaction effect for tree edit distance (p <.001) suggests that the combination of a summary description and a narrative style was particularly detrimental. These results suggest that detailed and technically-phrased inputs are most effective for accurately translating a user's creative intent into a functional DSL script, which may run counter to most players' natural speaking or writing styles.

\subsection{Auxiliary Findings}
Comparing our data generation methods, we found that scripts generated from grounded, procedurally-based inputs scored significantly higher on Creative Alignment (W=16968, p<.001, d=1.06), Instruction Following (W=16766, p<.001, d=1.00), and Emergence (W=24681, p<.001, d=0.55). Interestingly, the naturalistic inputs scored significantly higher on structural coherence (W=41737, p<.001, d=-0.58). We also found a clear trade-off between model performance and inference latency, with the fastest model (\textit{Gemma 3 4B}, M=3.35s) producing the lowest quality outputs and the slowest (\textit{Gemini 2.5 Flash}, M=12.2s) performing better.

\section{Human Pilot Study}\label{sec:human-pilot-study}
To complement and extend the automated evaluation, we conducted a small-scale pilot study with 6 human participants. The study was designed to capture a holistic player experience, which means there are significant methodological differences between the LLM judge's analysis of static code and the players' live play. Participants interacted with the prototype without knowledge of the underlying DSL and possessed genuine creative intent, whereas the LLM judge was provided with full documentation and could only infer intent from the provided input text.

Participants rated the generated mechanics on the same 1-5 scales for Creative Alignment, Instruction Following, and Emergence as the LLM judge. Ratings were generally high (median scores of 4, 4, and 3, respectively). We found a weak but statistically significant positive correlation between human and LLM ratings for Creative Alignment ($\rho$=.37, p<.01) and Instruction Following ($\rho$=.34, p<.01). This suggests that the programmatic qualities assessed by the LLM have a measurable, though modest, overlap with the subjective experience of a human player.

In contrast, no significant correlation was found for Emergence, implying that a player's subjective feeling of surprise is distinct from the judge model’s technical measure of novelty. Qualitative feedback showed that participants were often more lenient than the LLM judge, expressing enjoyment and surprise at the creative outcomes in the game, sometimes despite subpar interpretations of their inputs. However, some also described difficulty in determining the underlying capabilities of the system and what kind of requests they could expect it to handle. These findings indicate that while the LLM judge is a useful proxy for output quality, it cannot fully capture the subjective, interactive nature of human play.

\section{Discussion}

Our results indicate that the proposed LLM-ECS architecture is a promising method for translating natural language into executable game mechanics; Using an intermediate DSL proved effective for maintaining system stability by constraining the LLM to generate structured data. The architecture does not entirely eliminate the authoring burden, but changes the type of work necessary; developers must move from creating exponential content (e.g., every possible spell) to designing an expressive underlying system (the DSL and its components).

The choice of LLM was the most significant predictor of quality. Our findings show that a certain competence threshold, met by large state-of-the-art models, is required to handle the creative and logical reasoning for a novel DSL without fine-tuning, while the best prompting strategy depended on the structure of the target DSL. The procedural automata DSL, with its strict ordering, required few-shot examples to generate coherent scripts, whereas the more forgiving compositional spell DSL benefited more from a Chain-of-Thought reasoning step. This suggests that an important design consideration is that the nature of the task and structure of the DSL should inform the prompting strategy.

\paragraph{Potential for Emergent Gameplay}

One of the primary goals of this work was to develop a system that could enable dynamic and emergent gameplay, and a finding after implementation is that the nature of this emergence is closely tied to the level of abstraction that the LLM is permitted to operate. In the ECS-driven Battle Mode, novelty arose from surprising combinations of high-level, pre-authored capabilities. Instead of eliminating the historical authoring burden problem entirely, the developer's task moves from scripting every possible outcome to engineering a flexible and expressive set of foundational components. The resulting gameplay, while novel, is therefore constrained by the creative limits of that initial ECS component design, which requires consideration.

In contrast, Alchemy Mode demonstrated a more classic, ``bottom-up'' form of emergent gameplay by tasking the LLM with generating simple, local rules for material interactions from first principles. This approach produced complex and unpredictable large-scale patterns that were not present in the system's low-level parts. The relative success of this mode was likely driven by two key factors: a more granular, low-level DSL, and the extensive contextual grounding the model received on the simulation's current state. This implies that to encourage more complex, systemic emergence, a system must provide the model with enough detailed context and control over the most foundational building blocks of the simulation. This would shift designer's role from content creation into architecting a sufficiently expressive and robust environment, where the player can construct new high level rules through the LLM system.

\paragraph{Input Style \& Translation Fidelity}

The finding that the system performs best with precise, technical language presents a challenge for an interface designed to be intuitive; the system most accurately matched the proxy for creative intent when given these structured descriptions, suggesting that users may need to adopt a more technical phrasing to achieve predictable results. For creative applications, this is not necessarily a flaw; misinterpretations can lead to surprising and enjoyable outcomes, framing the LLM interaction as a creative, experimental process that can yield unexpected results. However, for scientific or other high-stakes applications requiring precision, this ambiguity presents a significant barrier. This suggests several directions for future HCI research, such as developing adaptive interfaces that include player feedback or systems that can outline plans or request clarification when faced with ambiguous commands.

\section{Conclusion}

This research introduces the LLM-ECS pattern as a reliable template for building safer, language-driven interactive systems. We provide a quantitative comparison of LLM performance on simultaneously creative and logical generation tasks and an analysis of how prompting strategies can be tailored to compositional versus procedural DSLs. Future work could focus on adaptive and multi-modal interfaces that teach users more effective interaction styles. A more ambitious direction would be to have the LLM determine the actual outcomes of gameplay on a semantic or logical basis, and provide interoperable hooks for other game systems to ground these outcomes in the virtual world. For game development, the immediate next step is to integrate this architecture into a full game with player progression, while a longer-term goal is to enable the LLM to alter more fundamental game systems, a significant step toward solving the historical authoring burden in interactive media.

\section*{Limitations}

The primary limitation of this work is the use of an LLM judge for qualitative evaluation. While validated, it is not a perfect substitute for human assessment, is subject to potential self-enhancement bias, and notably struggled to reliably evaluate the structural coherence of the procedural automata scripts. A methodological disconnect still remains, where a script can appear well-reasoned in code yet result in a poor gameplay experience, or vice versa. Therefore, the judge ratings are only interpreted on a relative basis between experimental groups, and cannot speak to broader success in a live game. The small-scale human pilot study, while encouraging, may not generalize to a broader player base.

Our experiments also relied on synthetic data. Inputs for the bidirectional experiment were indirectly grounded to the DSL's feature space, likely simplifying the translation task compared to ambiguous, authentic user input. Additionally, the system was only tested with ``good faith'' (at least moderately aligned with design goals) inputs in the automated evaluation; a production system would require more comprehensive input sanitization beyond the simple ``fizzle out'' spell failure fallback used here to handle adversarial or nonsensical prompts. Finally, these results represent a narrow snapshot of a rapidly evolving field; Using state-of-the-art models whose architectures are proprietary and frequently updated poses a significant challenge to the reproducibility of this work.

\section*{Ethical Considerations}

The human pilot study in Section \ref{sec:human-pilot-study} has been approved by the FESE Engineering Mathematics and Physical Sciences Ethics Committee at the University of Exeter.

While our application in a recreational game is low-risk, deploying a similar architecture in safety-critical domains like robotics would require more substantial controls to handle misaligned LLM outputs. While modern LLM providers have robust safety features, there remains a risk of insensitive or offensive content being generated. However, since the model is primarily producing DSL code for this prototype, the negative impact of unsafe outputs is minimized. Additionally, the prototype's reliance on user-provided, external API keys introduces data privacy and ownership concerns, as user data becomes subject to the terms of third-party model providers. This may be mitigated by opting out the human review and data sharing features in those providers. For many sensitive applications, privately-hosted models would be preferable.

\bibliography{custom}

\appendix

\section{Abridged Prompts}
\label{sec:appendixC}

This appendix contains the abridged, plain-text versions of the three main prompts used in our experiments. To adhere to formatting constraints, lengthy definitions of components and evaluation rubrics were replaced with descriptive tags, while all instructional text remains unaltered where possible. The full unabridged prompts are available in our GitHub repository\footnote{\url{https://github.com/austin-the-drake/real-time-world-crafting-wordplay-demo}}.

\subsection{Spell Scripting Prompt}\label{app:spell-prompt}
The first prompt, shown in Listing \ref{lst:prompt_spell}, configures the LLM for \textit{Latent Space's} Battle Mode. Its task is to translate a user's natural language description into a valid JSON object that conforms to a DSL, designed with a compositional, unordered structure where spells are defined as collections of components. This approach, which maps directly to the underlying ECS architecture, was found to be more forgiving for LLMs than strictly ordered logic.

\begin{lstlisting}[caption={Abridged Spell Scripting Prompt.},label={lst:prompt_spell},breaklines=true, basicstyle=\small\ttfamily, language={}]
You are an AI game design assistant. Your task is to generate a JSON object that defines a magical spell using a component-based system. The spell will be based on a provided description.

Overall Goal:
Create a single JSON object. The root of this object must always contain the key: "components". The value of "components" must be an array of individual component objects. The top-level spell may optionally contain the key: "count" when strictly appropriate for multi-cast. The top-level spell must finally contain the key: "friendlyName" containing a creative 2-3 word name for future reference.

Strict Output Requirements:
- The entire response must be a single, valid JSON object.
- Do not include any explanatory text, markdown formatting, or anything outside this single JSON object.
- Use only the componentTypes and their associated properties as defined below.
- For fields with "Possible Options" or specific enumerated values, you must choose a value from the provided list(s). Do not invent new string values for these fields.
- If a property is optional and not relevant to the spell concept, omit it.
- Properties of a spell are never automatically inherited by sub-spells; they must be repeated as necessary when using triggers.
- Always think creatively, and always consider whether to add sub-spells or physical manifestations when it could strengthen the concept.
- Numerical values should be sensible for a game context and fall within suggested ranges if provided.

Component Definitions:

A spell is defined by an array under the "components" key. Each object in this array is a component.

I. Spell Class Components (Choose exactly one as the primary form of the spell or sub-spell. Using more than one requires that they be placed in different nested triggers.):
- projectile: Defines the spell as a projectile.
  <... 7 more 'Spell Class' components were defined here (e.g., wallCrawl, aoe, shield, manifestation), each with specific properties and descriptions. Full list omitted for brevity. ...>

II. General Spell Property Components (Add as needed):
- element: Assigns a magical element from a predefined list.
  <... 3 more 'General Spell Property' components were defined here (e.g., color, spawnAngle, manaCost), each with specific properties. Full list omitted for brevity. ...>

III. Behaviour Modifier Components (Can be stacked; primarily affect projectile or wallCrawl):
- homing: Causes the spell to seek enemies.
  <... 2 more 'Behaviour Modifier' components were defined here (boomerang, controllable), each with specific properties. Full list omitted for brevity. ...>

IV. Trigger Components (Define sub-spells or effects triggered by conditions):
- timerTrigger: Executes a sub-spell after a set time.
  <... 3 more 'Trigger' components were defined here (buttonTrigger, impactTrigger, deathTrigger), each with specific properties to define a sub-spell payload. Full list omitted for brevity. ...>

<Few-shot examples may be provided here>

Your Task:
Generate the JSON object containing a "components" array for the spell concept provided below. Ensure all your choices and values adhere to the definitions, constraints, and suggested ranges listed above. The user's description of a magical spell will be provided first, followed by a list of all magical elements available in your toolbox.

<Dynamic context here>
<Player input here>
\end{lstlisting}

\subsection{Automata Scripting Prompt}\label{app:automata-prompt}
The second prompt (Listing \ref{lst:prompt_automata}) is tailored to the cellular automata sandbox. It instructs the LLM to generate a behavior script using a node-based DSL with a strict, procedural structure. Unlike the spell DSL, the sequence of conditional checks and actions in this language is highly important to the final outcome.

\begin{lstlisting}[caption={Abridged Automata Scripting Prompt.},label={lst:prompt_automata},breaklines=true, basicstyle=\small\ttfamily, language={}]
You are a game design assistant specialising in cellular automata. Your task is to generate a single, valid JSON object that defines a behavior script based on the user's description. The output must be a single JSON object containing three root keys: "name" (a creative, one-word, lowercase string), "color_hex" (a hex string like "#RRGGBB"), and "behavior" (a struct containing the "actions" array).

The entire response must be a single, valid JSON object. Do not include any explanatory text, markdown formatting, or anything outside of this object. Adhere strictly to the node definitions provided.

Important Notes:
- The term "direction" below is a placeholder for one of ["north", "northeast", "east", ...].
- "self" is a valid keyword for a particle's own cell.
- Actions are processed sequentially.
- The do_swap node is special: it moves the cell, then can immediately run a nested actions list from the cell's new location. [...]
- If you are asked to update an existing cell type, do not alter its name field, or else it may break references.

Available Node Types (for the actions array inside the behavior struct)

I. Wrapper / Modifier Nodes (These contain other nodes):
- in_rand_rotation: Executes nested actions in one random of 8 directions.
  <... 2 more 'Wrapper / Modifier' nodes were defined here (in_rand_mirror, in_rand_flip). Full list omitted for brevity. ...>

II. Conditional Nodes (These check a condition and then run nested actions):
- if_neighbor_is: Checks if a neighbor is one of the types in options.
  <... 4 more 'Conditional' nodes were defined here (e.g., if_alpha, if_neighbor_count, if_chance). Full list omitted for brevity. ...>

III. Executor / Action Nodes (These perform an action and are usually the innermost nodes):
- do_swap: Swaps position with a neighbor, then can immediately run a nested actions list from the cell's new location. This entire operation ends the cell's turn.
  <... 4 more 'Executor / Action' nodes were defined here (e.g., do_set_type, do_spawn, do_copy_alpha). Full list omitted for brevity. ...>

<Few-shot examples may be provided here>

Your Task:
Be creative and thorough. [...] Now, generate the complete JSON object for the following element. You will be provided with a user input [...], and a list of all the existing materials and their respective behaviors.

<Dynamic context here>
<Player input here>
\end{lstlisting}

\subsection{Judge Model Prompt}\label{app:judge-prompt}
The third prompt (Listing \ref{lst:prompt_judge}) configures a separate LLM to act as an automated judge. This approach was justified by established literature on using LLMs for larger-scale qualitative assessments. To improve reliability and mitigate known issues like leniency bias, the prompt uses several strategies, including instructing the model to produce a rationale before its final score (a form of Chain-of-Thought) and using a capable state-of-the-art model, in line with recommended guidelines.

\begin{lstlisting}[caption={Abridged Judge Prompt.},label={lst:prompt_judge},breaklines=true, basicstyle=\small\ttfamily, language={}]
Your primary role is to evaluate a JSON-based (spell/automata) script. You will act as a strict technical reviewer. This is a difficult logical and creative task, and most scripts will not achieve a perfect score.

This prompt provides all the information you need in a structured format. First, you will be presented with the task rules [...]. Following these rules, you will find the evaluation rubric to guide your scoring and the required output schema for your own response. Finally, you will see the full task [...].

Please begin by carefully reviewing the documentation below.

<Documentation from the original game prompt was provided to the model here.>

You will evaluate the generated script based on the following four axes. This is a challenging task and you should evaluate critically. Score each axis from 1 (poor) to 5 (excellent).

Creative Alignment (1-5):
(This scale measures how well the code matched the core concept or "theme" of the user's description, with an emphasis on presentation.)
- 5 (Perfect Match): The code perfectly captured the core concept and all thematic details, feeling exactly like what was imagined.
- 1 (No Connection): The code's core function was the opposite of what was requested.
  <... Detailed descriptions for scores 2-4 on this axis were also provided. ...>

Instructional Precision (1-5):
(This scale measures how well the code obeyed any specific, direct commands or constraints the user included, with an emphasis on logic.)
- 5 (Followed All Instructions Perfectly): Every single explicit instruction and constraint was implemented precisely as requested.
- 1 (Ignored All Instructions): The code failed to follow multiple explicit instructions.
  <... Detailed descriptions for scores 2-4 on this axis were also provided. ...>

Emergence (1-5):
(This scale measures if the code's behaviour went above and beyond the user's request in a positive way.)
- 5 (Delightful Surprise): The code added a significant, unrequested feature that was a perfect creative fit, making the entire concept much better and more interesting.
- 1 (Boring/Literal): The code was a completely literal and uninspired translation of the prompt with no creative flair whatsoever.
  <... Detailed descriptions for scores 2-4 on this axis were also provided. ...>

Structural & Logical Coherence (1-5):
(This scale evaluates the syntactic and logical validity of the generated spell.) This is informed by the "Algorithmic Pre-Check". If the pre-check failed, this score cannot be higher than 2.
- 5: The entire response is a single, valid JSON object with perfect data types and structure, and it is completely free of any logical contradictions defined in the component rules.
- 3: Valid JSON, but with some data type errors or logical conflicts forbidden by the rules (e.g., a manifestation component having a deathTrigger).
- 1: Invalid JSON or major structural errors (e.g., components is not a list).

<Calibration examples were provided to the model here.>

You must respond with a single, valid JSON object and nothing else. Remember that these responses must be graded strictly. The majority of responses should achieve an average of 2-4. 4s and 5s should be reserved for truly exceptional responses [...].

The root object of your response must contain two keys, in this exact order: "rationales" and "scores".
- "rationales": An object containing a brief, one-sentence text explanation for each of the four scores. [...]
- "scores": An object containing a numerical score (1-5) for each of the four rubric axes. [...]
\end{lstlisting}

\section{Full NL-to-DSL ANOVA Results}
\label{sec:appendixA}

\begin{table*}[t]
\centering
\footnotesize
\caption{ANOVA Results for the NL-to-DSL Experiment}
\label{tab:exp1_anova}
\begin{tabular*}{\textwidth}{@{\extracolsep{\fill}}llcccccc@{}}
\toprule
& & \multicolumn{3}{c}{\textbf{Magical Spells}} & \multicolumn{3}{c}{\textbf{Cellular Automata}} \\
\cmidrule(lr){3-5} \cmidrule(lr){6-8}
\textbf{Outcome} & \textbf{Predictor} & \textbf{F} & \textbf{p} & \textbf{$\eta_G^2$} & \textbf{F} & \textbf{p} & \textbf{$\eta_G^2$} \\
\midrule

Creative Alignment & Model (M) & \textbf{58.43} & \textbf{$<$.001\ddag} & \textbf{.12} & \textbf{117.88} & \textbf{$<$.001\ddag} & \textbf{.35} \\
& Shot Strategy (S) & 0.53 & .581 & .00 & \textbf{21.13} & \textbf{$<$.001\ddag} & \textbf{.03} \\
& Planning (P) & \textbf{9.04} & \textbf{.004\dag} & \textbf{.01} & \textbf{15.85} & \textbf{$<$.001\ddag} & \textbf{.01} \\
& M $\times$ S & \textbf{2.31} & \textbf{.040*} & \textbf{.01} & 1.14 & .340 & .00 \\
& M $\times$ P & 2.33 & .080 & .00 & 0.78 & .499 & .00 \\
& S $\times$ P & 0.29 & .744 & .00 & 0.91 & .400 & .00 \\
& M $\times$ S $\times$ P & 0.95 & .447 & .00 & 1.05 & .386 & .00 \\
\addlinespace

Instruction Following & Model (M) & \textbf{86.86} & \textbf{$<$.001\ddag} & \textbf{.14} & \textbf{115.53} & \textbf{$<$.001\ddag} & \textbf{.31} \\
& Shot Strategy (S) & 1.47 & .235 & .00 & \textbf{23.75} & \textbf{$<$.001\ddag} & \textbf{.03} \\
& Planning (P) & 2.78 & .102 & .00 & \textbf{7.98} & \textbf{.007\dag} & \textbf{.01} \\
& M $\times$ S & \textbf{2.54} & \textbf{.024*} & \textbf{.01} & 1.47 & .199 & .01 \\
& M $\times$ P & \textbf{2.77} & \textbf{.044*} & \textbf{.01} & 2.14 & .107 & .00 \\
& S $\times$ P & 1.91 & .153 & .00 & 1.17 & .316 & .00 \\
& M $\times$ S $\times$ P &1.04 & .398 & .00 & 0.77 & .571 & .00 \\
\addlinespace

Emergence & Model (M) & \textbf{35.63} & \textbf{$<$.001\ddag} & \textbf{.10} & \textbf{128.81} & \textbf{$<$.001\ddag} & \textbf{.33} \\
& Shot Strategy (S) & 1.81 & .170 & .00 & \textbf{16.08} & \textbf{$<$.001\ddag} & \textbf{.02} \\
& Planning (P) & \textbf{5.78} & \textbf{.020*} & \textbf{.01} & \textbf{17.57} & \textbf{$<$.001\ddag} & \textbf{.01} \\
& M $\times$ S & \textbf{2.66} & \textbf{.021*} & \textbf{.01} & 1.62 & .156 & .01 \\
& M $\times$ P & 0.81 & .487 & .00 & 1.00 & .386 & .00 \\
& S $\times$ P & \textbf{3.63} & \textbf{.032*} & \textbf{.00} & 0.23 & .790 & .00 \\
& M $\times$ S $\times$ P & \textbf{2.66} & \textbf{.020*} & \textbf{.01} & 0.83 & .521 & .00 \\
\addlinespace

Structural Coherence & Model (M) & \textbf{192.28} & \textbf{$<$.001\ddag} & \textbf{.41} & \textbf{18.81} & \textbf{$<$.001\ddag} & \textbf{.05} \\
& Shot Strategy (S) & 2.88 & .062 & .00 & \textbf{22.32} & \textbf{$<$.001\ddag} & \textbf{.05} \\
& Planning (P) & 0.38 & .540 & .00 & 1.55 & .220 & .00 \\
& M $\times$ S & \textbf{2.56} & \textbf{.031*} & \textbf{.01} & \textbf{7.20} & \textbf{$<$.001\ddag} & \textbf{.03} \\
& M $\times$ P & \textbf{6.31} & \textbf{.001\dag} & \textbf{.02} & 0.78 & .486 & .00 \\
& S $\times$ P & 1.48 & .233 & .00 & 0.01 & .968 & .00 \\
& M $\times$ S $\times$ P & 0.79 & .558 & .00 & 0.91 & .450 & .00 \\
\bottomrule

\multicolumn{8}{l}{\textit{Note.} $\eta_G^2$ = generalized eta-squared. Significance codes: * $p < .05$, \dag\ $p < .01$, \ddag\ $p < .001$} \\
\end{tabular*}
\end{table*}

\begin{figure*}[t]
\centering	
\includegraphics[width=0.9\linewidth]{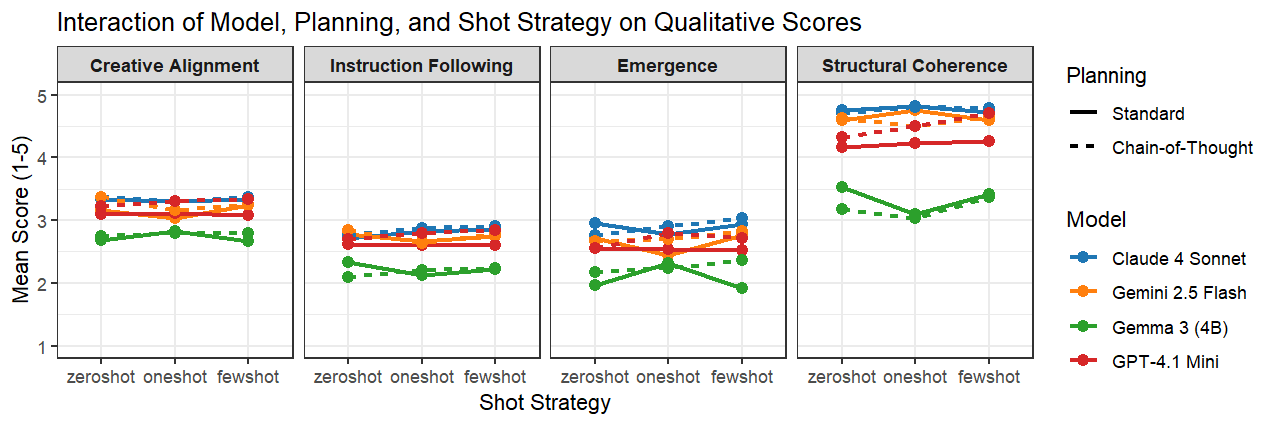}
\caption{Interaction plots for the NL-to-DSL ANOVA for magical spell DSL scripts.
}\label{fig:exp1_interactions}
\end{figure*}

The full ANOVA results for the NL-to-DSL translation experiment are presented in Table \ref{tab:exp1_anova}. The analysis shows that the choice of LLM was the most significant predictor of output quality across all four qualitative measures for both the compositional spell DSL and the procedural automata DSL (p<.001). The effects of prompting strategies were highly task-dependent; Chain-of-Thought (CoT) prompting generally improved creative alignment and emergence, while the inclusion of few-shot examples was only significant for the more logically complex cellular automata task.

Figure \ref{fig:exp1_interactions} visualizes these results for the magical spell DSL, demonstrating the significant interaction effects found between the model, shot strategy, and prompting technique. This illustrates that the optimal prompting strategy is highly dependent on the specific model being used, as the impact of CoT and few-shot examples was not uniform across all models.

\section{LLM Judge Validation}
\label{sec:appendixB}

\begin{figure*}[h!]
	\centering	
	\includegraphics[width=0.9\linewidth]{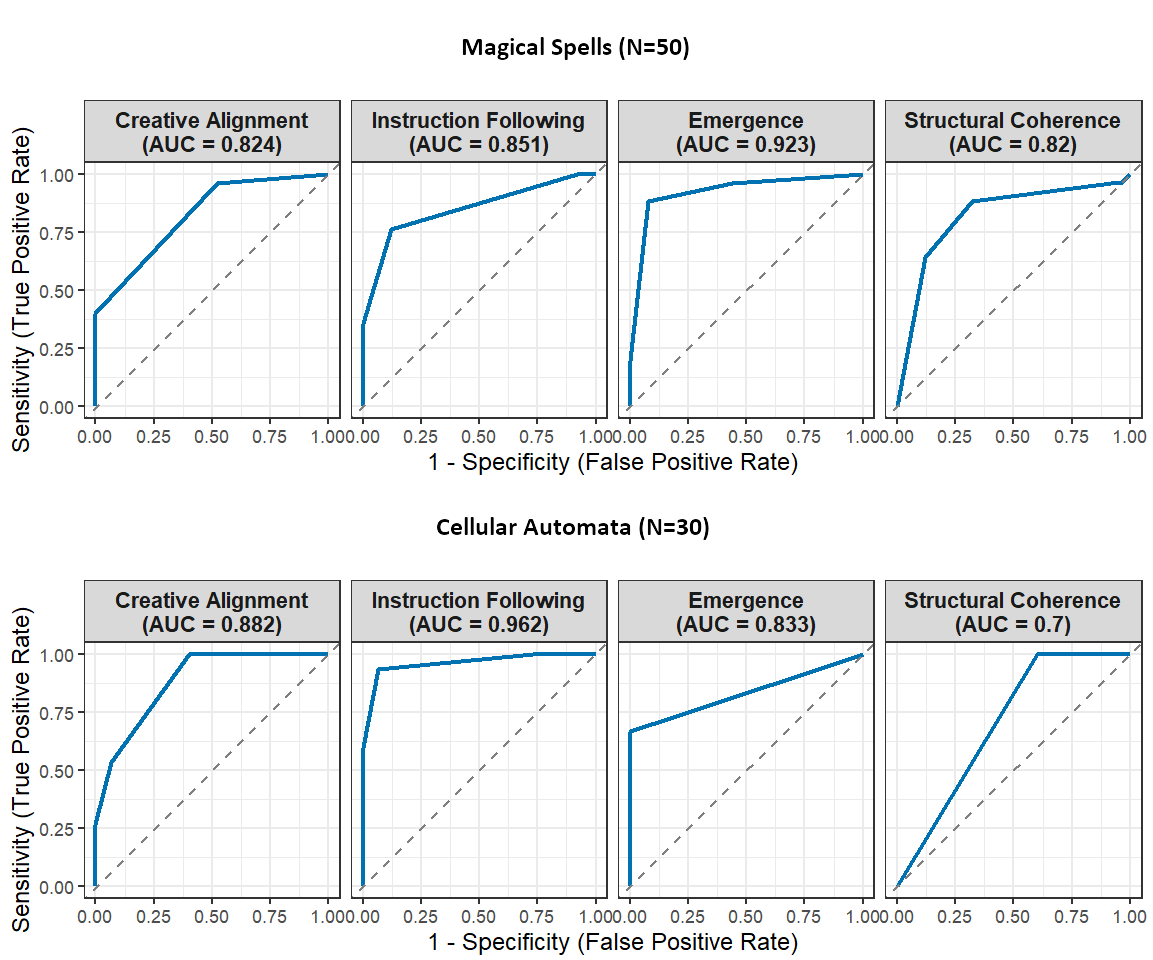}
	\caption{ROC plots of \textit{GPT-4.1} performance in discriminating between manually-authored ``good'' and ``bad'' DSL scripts. Performance was evaluated separately for the 4 Likert scales on the ground-truth data.
    }\label{fig:rocs_gpt}
\end{figure*}

\begin{table*}[h!]
\centering
{\footnotesize
\caption{Inter-Rater Reliability Metrics Between \textit{GPT-4.1} and \textit{Gemini 2.5 Pro}}
\label{tab:irr_metrics}
\begin{tabular*}{\textwidth}{@{\extracolsep{\fill}}lcccccc@{}}
\toprule
& \multicolumn{3}{c}{\textbf{Magical Spells}} & \multicolumn{3}{c}{\textbf{Cellular Automata}} \\
\cmidrule(lr){2-4} \cmidrule(lr){5-7}
\textbf{Metric} & \textbf{Spearman's $\rho$} & \textbf{Weighted $\kappa$} & \textbf{ICC} & \textbf{Spearman's $\rho$} & \textbf{Weighted $\kappa$} & \textbf{ICC} \\
\midrule
Creative Alignment & .70 & .57 & .57 & .81 & .74 & .75 \\
Instruction Following & .68 & .57 & .57 & .79 & .75 & .75 \\
Emergence & .68 & .64 & .65 & .76 & .77 & .77 \\
Structural Coherence & .59 & .56 & .56 & .36 & .23 & .24 \\
\bottomrule

\end{tabular*}
}
\end{table*}

This appendix details the validation process for the LLM judge used in our experiments. Figure \ref{fig:rocs_gpt} presents the Receiver Operating Characteristic (ROC) curves, which visualize the performance of the \textit{GPT-4.1} judge in discriminating between expert-authored ``good'' (well-formed and effective) and ``bad'' (logically flawed) scripts from our handcrafted ground-truth dataset (N=80). The high Area Under the Curve (AUC) values across most criteria demonstrate the judge's strong alignment with a human-defined standard of quality.

To further ensure consistency and mitigate single-model bias, we calculated the inter-rater reliability between the primary judge and a second, equivalently-capable model, \textit{Gemini 2.5 Pro}. Table \ref{tab:irr_metrics} displays these metrics, including Spearman's $\rho$, weighted Cohen's Kappa ($\kappa_w$), and the Intraclass Correlation Coefficient (ICC). The results show moderate to substantial agreement across most criteria, indicating that the evaluation scales are objective enough to be scored consistently by different state-of-the-art models. A notable exception was lower agreement on the Structural Coherence scale for the complex cellular automata scripts, suggesting this type of logical evaluation remains challenging even for capable models.

\end{document}